\documentclass[aps,pra,twocolumn,groupedaddress,reprint]{revtex4}
\usepackage{graphicx}
\usepackage{setspace}
\usepackage{amsmath}
\usepackage{amssymb}
\usepackage{color}
\usepackage{array}
\usepackage{subfigure}
\usepackage{hyperref}
\usepackage{float}
\usepackage{lipsum}
\usepackage{amsfonts}
\usepackage{mathrsfs}
\usepackage{dcolumn}
\usepackage{bm}
\usepackage{bbm}
\usepackage{graphicx}

\begin{document}
\title{Effect of source tampering in the security of quantum cryptography}
\author{Shi-Hai Sun$^1$}
\altaffiliation{shsun@nudt.edu.cn}
\author{Feihu Xu$^{2,4}$}
\altaffiliation{tigerfeihuxu@gmail.com}
\author{Mu-Sheng Jiang$^1$, Xiang-Chun Ma$^1$, Hoi-Kwong Lo$^2$}
\altaffiliation{hklo@ece.utoronto.ca}
\author{ Lin-Mei Liang$^{1,3}$}
\altaffiliation{nmliang@nudt.edu.cn}

\affiliation{$^1$ College of Science, National University of Defense
Technology, Changsha 410073, China\\
$^2$ Center for Quantum Information and Quantum Control, Department of  Electrical and Computer Enginnering and Department of Physics, University of Toronto, Toronto, Ontario, M5S 3G4, Canada\\
$^3$State Key Laboratory of High Performance Computing, National University of Defense Technology, Changsha 410073, China\\
$^4$ Research Laboratory of Electronics, Massachusetts Institute of Technology, 77 Massachusetts Avenue, Cambridge, Massachusetts 02139, USA}

\date{\today}
\begin{abstract}
The security of source has become an increasingly important issue in quantum cryptography. Based on the framework of measurement-device-independent quantum-key-distribution (MDI-QKD), the source becomes the only region exploitable by a potential eavesdropper (Eve). Phase randomization is a cornerstone assumption in most discrete-variable (DV-) quantum communication protocols (e.g., QKD, quantum coin tossing, weak coherent state blind quantum computing, and so on), and the violation of such an assumption is thus fatal to the security of those protocols. In this paper, we show a simple quantum hacking strategy, with commercial and homemade pulsed lasers, by Eve that allows her to actively tamper with the source and violate such an assumption, without leaving a trace afterwards. Furthermore, our attack may also be valid for continuous-variable (CV-) QKD, which is another main class of QKD protocol, since, excepting the phase random assumption, other parameters (e.g., intensity) could also be changed, which directly determine the security of CV-QKD.
\end{abstract}


\maketitle
\section{Introduction}
Quantum key distribution (QKD) \cite{BB84} allows two remote parties to share an unconditional secret key, which has been proven in theory \cite{Lo99,Shor00,GLLP04} and demonstrated in experiment \cite{Wang12}. However, the imperfections of practical devices will compromise the security of QKD systems \cite{Zhao08,Xu10,Gerhardt11,Lydersen10,Bugge14,Sun11,Jain11,Weier11,Ma13}. So far, three main approaches have been proposed to bridge the gap between theory and practice. The first one is to close specific loopholes of devices with security patches  \cite{Yuan10}, but it could not close potential and unnoticed loopholes. The second one is device-independent (DI-) QKD \cite{Mayers98,Acin07,Pironio09}. By testing Bells inequality in a loophole-free setting, security could be obtained without detailed information about the implementation devices. But DI-QKD is impractical because an almost perfect single photon detector (SPD) is required, and even so the secret key rate is limited \cite{Curty11,Gisin10}. The third approach is to remove as many device loopholes and assumptions as possible by either modifying the QKD protocol or refining the security proof. One of the best results with this approach is measurement-device-independent (MDI-) QKD \cite{Lo12}, which can remove all detector loopholes. Since the detection system is widely regarded as the Achilles' heel of QKD \cite{Zhao08,Gerhardt11,Lydersen10,Bugge14}, MDI-QKD is of great importance. Indeed, recently, MDI-QKD has been demonstrated both in the laboratory and in the field \cite{MDIExp}.

Based on the framework of MDI-QKD, the source becomes the final battlefield for the legitimate parties and Eve. And the major flaw of the source is that a semiconductor laser diode (S-LD), which generates a weak coherent state, is normally used as a single photon source in most commercial and research QKD systems \cite{Wang12,MDIExp}. The security of MDI-QKD as well as BB84 based on S-LD has been proven with decoy state \cite{Decoystate}. Hence, it has been convinced that if the source can be well characterized (for example the source flaws could be taken care of with the loss-tolerant QKD protocol \cite{Tamaki14}), perfect security can still be obtained.

Generally speaking, there are two main classes of QKD protocols, one is discrete-variable (DV-) QKD (including BB84, decoy state BB84, MDI-QKD, Scarani-Acin-Ribordy-Gisin (SARG04) \cite{Scarani04}, differential phase shift  (DPS) \cite{Takesue05}, and so on) , and the other one is continuous-variable (CV-) QKD \cite{cvQKD}. In most DV- quantum communication protocols (e.g., DV-QKD, quantum coin tossing (QCT) \cite{Pappa14}, weak coherent state blind quantum computing (BQC) \cite{Dunjko12}), the phase randomization is a cornerstone assumption. By assuming that the overall phase is uniformly distributed from 0 to $2\pi$ (in fact, discrete randomization with finite points, e.g., 10, is sufficient to guarantee QKD security \cite{Cao14}), a coherent state with intensity $|\alpha|^2$ is reduced into a classical mixture state, that is $\rho=\int_0^{2\pi}\frac{d\theta}{2\pi}|\alpha e^{i\theta}\rangle\langle\alpha e^{i\theta}|=\sum_{n=0}^\infty \frac{e^{-|\alpha|^2}|\alpha|^{2n}}{n!}|n\rangle\langle n|.$ Then it allows one to apply classical statistics theory to analyze quantum mechanics. Note that although the security of QKD with nonrandom phase had been proven \cite{Lo07}, the performance is very limited in distance and key rate.

In this paper, however, we demonstrate a simple quantum hacking strategy, with both a commercial and homemade pulsed laser based on S-LD, that allows Eve to actively violate the phase randomization assumption, without leaving a trace afterwards. Thus it is effective for most of DV- quantum communication protocols. Our attack may also be effective for CV-QKD, since other parameters of the source (e.g., intensity) could also be changed. For example, it had been proven that the local oscillator fluctuation will compromise the security of CV-QKD \cite{Ma13}. Since S-LDs are widely used in most quantum information protocols (e.g., DV-QKD, CV-QKD, QCT, BQC, and so on), and the security of these protocols is closely related to S-LD's parameters \cite{GLLP04}, our work constitutes an important step towards secure quantum information processing.

Our attack differs from previous attacks \cite{Zhao08,Xu10,Gerhardt11,Lydersen10,Bugge14,Sun11,Jain11,Weier11,Ma13}. First, in our attack, Eve actively violate some basic assumptions required in the security proof by tampering with an initial perfect source. Second, unlike the laser damage attack \cite{Bugge14} in which Eve also actively creates loopholes for a perfect SPD, the loopholes created by our attack are temporary, this makes our attack impossible for Alice and Bob to detect during the off-time of the QKD system. Third, our attack also differs from the Trojan-horse attack \cite{Gisin06,Jain14}. In our attack, Eve directly break some basic assumptions of QKD protocols, whereas in the Trojan horse attack, back-reflected light is measured to analyze Alice's information. And as the best we know, the Trojan horse attack is invalid for Alice with multi-lasers \cite{Schmitt07}, but our attack remains applicable to such systems. Fourth and most importantly, our attack targets the source instead of SPD. This makes our attack a serious threat for most quantum information protocols (not only QKD, but also QCT and BQC).

Here we emphasize that the phase randomization is a cornerstone assumption in the security of many quantum communication protocols including QKD, QCT and BQC. It is important for not only weak coherent pulse based protocols, but also, for instance, parametric down conversion based protocols \cite{QWang08}. And continuous or discrete phase randomization is also crucial for the loss-tolerant protocol \cite{Tamaki14}. In fact, without the phase randomization, the performance of a quantum communication protocol will be dramatically reduced in distance and key rate \cite{Lo07}.  However, we demonstrate experimentally in a clear manner how easy it is for Eve to violate such a fundamental assumption in a practical setting. Thus our work is very generality for most of quantum information processing protocols. It works for most DV-QKD, with various encoding schemes (polarization, phase and time-bin) and various kinds of lasers (pulsed laser and continuous wave (cw) laser). It is also possibly a serious threat for CV-QKD and other quantum information processing protocols (such as QCT and BQC).

The basic principle of our attack is as follows. In the inter-driven mode, the semiconductor medium of the S-LD is excited from loss to gain by each driving current pulse. A laser pulse is generated from \emph{seed} photons originating from spontaneous emission. The phase of the laser pulse is determined by the seed photons. Since the phase of the seed photons is random, the phase of each laser pulse is random inherently \cite{Williams10,Xu12,Yuan14,Kobayashi14}. However, if a certain number of photons are injected from an external source into the semiconductor medium, these photons will also be amplified to generate laser pulses. Consequently, the seed photons consist of two parts: one from spontaneous emission and the other part from the external source. Both parts will affect the phase of the resulting laser pulse. If the injected photons greatly outnumber the photons from spontaneous emission, the phase of the output laser pulse is largely determined by the phase of the injected photons. Therefore, Eve can control the phase of Alice's signal laser by illuminating the S-LD from an external `control source', and successfully violate the phase randomization assumption.

\begin{figure}
\scalebox{1}{\includegraphics[width=\columnwidth]{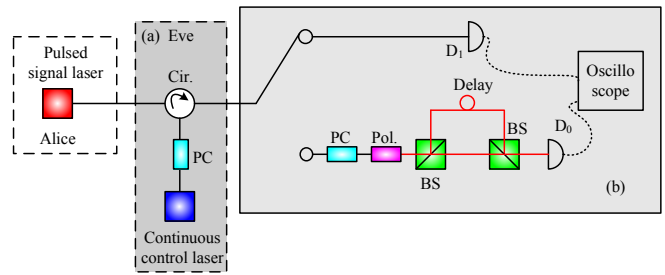}}
\caption{\label{fig:scheme}(Color online) Schematic setup of our experiment. Part(a) shows Eve's control devices, in which Eve uses a continuous wave (cw) laser to tamper with the parameters of Alice's pulsed signal laser. Part(b) shows the experimental setups to measure the parameters of Alice's signal pulses. The phase of adjacent pulse is measured by an unbalanced Mach-Zehnder interferometer [lower arm of part(b)]. And the waveform of Alice's signal pulse is directly measured with a photodiode [upper arm of part(b)]. The output of photodiodes ($D_0$ and $D_1$) are recorded with an oscilloscope. Cir.: circulator; PC: polarization controller; Pol.: polarizer; BS: beam splitter. Solid lines are optical fibers (single-mode fiber for black color and polarization-maintaining fiber for red color), and dashed lines are electrical lines. Here we consider Eve's control laser working at continuous wave (cw) mode. However, in later parts of this paper, we will consider the possibility that Eve modulates her control laser into short photon pulses. This can make it harder for Alice to detect Eve's attack.}
\end{figure}

\begin{figure}
\scalebox{1}{\includegraphics[width=\columnwidth]{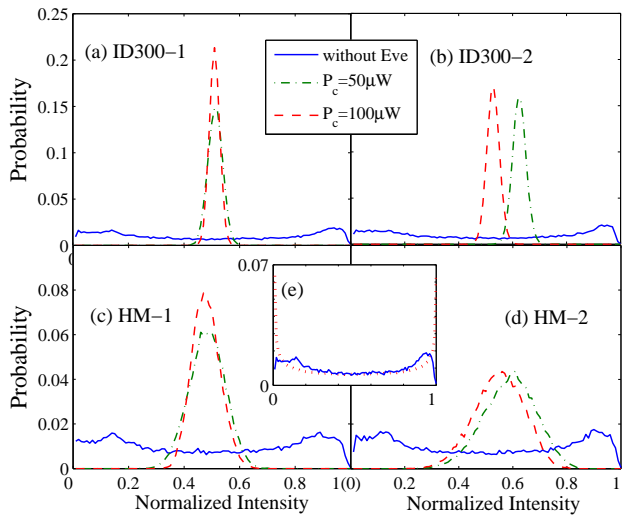}}
\caption{\label{fig:pro}(Color online) Experimental results for normalized intensity distribution of $V_P^s$. $P_c$ is the power of Eve's control laser. Parts(a)-(d) show the intensity distribution of four S-LDs with Eve's different control intensities. Part(e) shows the theoretical simulation (dashed line) of the probability distribution when the phase of each pulse follows a uniform distribution from 0 to $2\pi$, and the experimental results of ID300-1 (solid line) when Eve is absent. These results clearly show that when photons are injected into Alice's signal laser, the phase of the signal laser becomes correlated. Here $P_c$ is not minimized for Eve \cite{CommentFig2_2}, and a further experiment about the minimal power is discussed in the following text (see Fig.\ref{fig:pro_isolator}).}
\end{figure}

\section{Experiment and main results}
Figure \ref{fig:scheme} shows the schematic setup of our experiment. We test four sample S-LDs operating in inter-driven mode, two ID300 pulsed lasers from IdQuantique \cite{IDQ} (numbers ID300-1 and ID300-2), and two homemade pulsed lasers with S-LDs from Sunstar Communication Technology CoLtd (model: SDLP55HMBIFPN, numbers HM-1 and HM-2). To measure the phase relationship between adjacent pulses, an unbalanced Mach-Zehnder interferometer is used (see Fig.\ref{fig:scheme}(b)). The repetition rate of the signal laser is set to be 206.34MHz to match the delay of the interferometer. The output light is detected by a photodiode ($D_0$) with a bandwidth of 1 GHz, and the voltage of each pulse is recorded using an oscilloscope with bandwidth 33 GHz and sample rate 80 GHz (Agilent, model: DSOX93304Q).

Because the central frequency (with a finite linewidth) and polarization of the signal laser are unstable in experiment, Eve needs to carefully modulate the frequency and polarization of her control laser to match her control laser with Alice's signal laser. In our experiment, a tuning laser module (model: 81600B-201, Agilent) is used as Eve's control laser. Furthermore, in Fig.1 of the main text, we consider Eve's control laser working at cw mode. However, at the end of this paper, we consider the possibility that Eve modulates her control laser into short photon pulses. This can reduce Alice's ability to detect Eve's attack.

In theory, the output voltage after $D_0$ is
$V_P \propto [1+\cos(\Delta\phi+\theta_0)]/2$,
where $\Delta\phi$ is the phase difference between adjacent pulses, and $\theta_0$ is the inherent phase difference between the two paths of the interferometer. By passively controlling the interferometer with temperature controller and vibration isolator, we can stabilize the interferometer within about 2 minute. In the test we set the number of pulses to be 25791 in each experimental point of Fig.\ref{fig:pro} (In each experimental point of Fig.\ref{fig:pro}, we collect and store 10M data. Note that the repetition rate of the laser is 206.34 MHz and the sample rate of oscilloscope is 80GHz. The number of data is about (1/206.34MHz)/(1/80GHz)$\approx$ 388 in each pulse cycle. Thus the number of pulses is about $10M/388\approx 25791$.), and the time interval is about 0.125 ms ($25791/206.34MHz)$, which is much lower than the time scale of the interferometer. Thus we could set $V_P^s \propto [1+\sin(\Delta\phi)]/2$ for $\theta_0=\pi/2$.

\begin{figure}
\scalebox{1}{\includegraphics[width=\columnwidth]{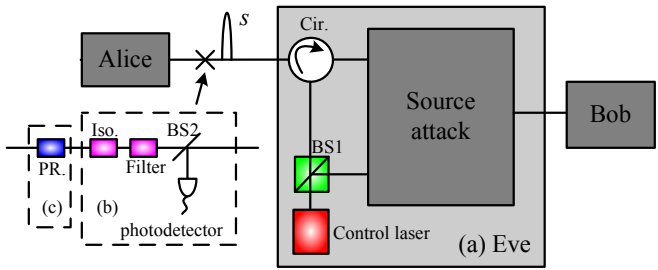}}
\caption{\label{fig:attack}(Color online) (a) Principle scheme to attack a complete QKD system by combining our attack with those of \cite{Sun12}. $s$ is Alice's quantum signal pulse. Eve splits her bright control pulse into two parts with a beam splitter (BS1), one part serves as control laser to tamper with the parameters of Alice's signal pulse, while the other part serves as phase reference for Eve to perform the source attack \cite{Sun12}. (b) A possible countermeasure for Alice to monitor our attack. Alice splits parts of the light with BS2 and monitors the power with a photodetector. The optical frequency filter is used to remove all wavelength-dependent flaw of Alice's source. The isolator (Iso.) is used to prevent light from entering Alice's lab from the quantum channel. (c) Active phase randomization scheme (PR.), which can guarantee the phase randomization assumption and partially reduce the risk of our attack, but it can not entirely remove our attack (see text for detail).}
\end{figure}

A uniform distribution of $\Delta\phi$ from 0 to $2\pi$ will produce a U-type intensity distribution, owing to the fact that the mapping from phase to intensity is non-linear, $V_P \propto \sin(\Delta\phi)$. Indeed when Eve is absent, the same distributions (solid lines of Fig.\ref{fig:pro}) are obtained in experiments with both ID-300 and the homemade pulsed laser. However, a bright light from Eve could correlate the phase of each pulse and violate the phase randomization assumption (dashed lines of Fig.\ref{fig:pro}). In fact, when photons are injected into Alice's signal laser, the intensity distribution of $V_P^s$ for both ID300 and the homemade signal laser becomes Gaussian. Consequently, various quantum hacking strategies can be applied to spy on the final key \cite{Sun12}. Figure \ref{fig:attack}(a) shows a schematic setup to attack a complete QKD system.

Theoretically speaking, Eve can perfectly control the phase of Alice's source, and then the intensity distribution should be a sharp line. However, owing to the following two main reasons, the measured intensity distribution in Fig.\ref{fig:pro} of the main text follows Gaussian distribution: (1) There exists phase noise in Eve's controlling laser, which follows Gaussian distribution. The measured intensity is the interference of adjacent pulses (the interval of adjacent pulses is about 5ns), thus the experimental results depends on the phase noise of Eve's control laser at different time. (2) The interference is imperfect, including the loss of two paths of the interferometer, the time jitter of optical pulse, and so on. Therefore, a practical Eve can't perfectly control the phase of Alice's source, and the phase noise decides how much information will be leaked to Eve. Furthermore, although the security of the BB84 protocol had been proven based on uniformed random phase from 0 to $2\pi$ \cite{GLLP04} and nonrandom phase \cite{Lo07}, but the key rate (or mutual information between Alice and Eve) is still unknown, if the phase of source follows Gaussian distribution or a general probability distribution, which will be studied in future.

Furthermore, we note that when the LD is operated in inter driven mode, the emitted pulses have random phase, and such phase noise had been used as a quantum random number generator by many groups \cite{Williams10,Xu12,Yuan14,Kobayashi14}. However, Fig.\ref{fig:pro}(e) of the main text does not prove that the phase of each pulse follows uniform distribution from $0$ to $2\pi$. In fact, if the phase is uniformly distributed from 0 to $\pi$, the same probability distribution could also be obtained. Thus, the phase randomization assumption must be carefully evaluated, particularly for a high-speed QKD system \cite{Kobayashi14}. Active phase randomization \cite{Zhao07} is a good countermeasure to guarantee the phase randomization assumption.

\begin{figure}
\scalebox{1}{\includegraphics[width=\columnwidth]{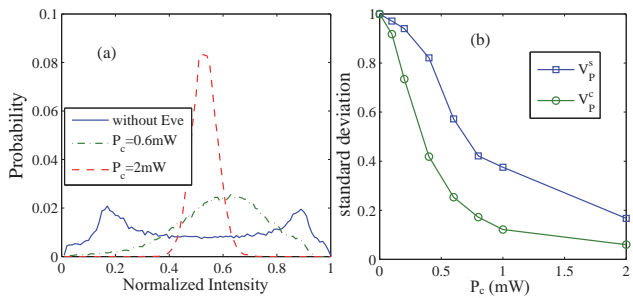}}
\caption{\label{fig:pro_isolator}(Color online) (a) Experimental results for $V_P^s$, when a 25dB isolator is placed after the signal laser ID300-1. (b) The standard deviation of $V_P^s \propto \sin(\Delta\phi)$ and $V_P^c \propto \cos(\Delta\phi)$ with different powers of control light. The standard deviation has been normalized by that of $P_c=0$. The experimental results clearly show that, even if a 25dB isolator is used by Alice, the intensity distribution is still Gaussian-type but not U-type when Eve uses a cw laser with a power of $0.6mW$, which means that Eve could still introduce a nonrandom phase in Alice's quantum signal. In the test, only a 25dB isolator is put after the output of Alice (the photodetector and the filter will be discussed later). Other setups used here are the same as those for Fig.\ref{fig:pro}.}
\end{figure}

\section{Countermeasure}
Figure \ref{fig:attack}(b) shows a possible countermeasure for Alice to monitor our attack. It includes three main devices, an isolator (Iso.), a filter and a photodetector. But these devices could not defeat our attack completely, if they are not carefully configured (see Appendix \ref{appendix_a} for details). (1) The isolator could not entirely stop Eve's photons due to its finite isolation (see Fig.\ref{fig:pro_isolator}), and other imperfections of practical isolators have been found in a recent paper \cite{Jain14}. (2) Since the wavelength of Eve's control laser is the same as that of Alice's signal laser in our attack, an optical frequency filter is also ineffective. (3) Both optical power meter and classical photodetector could be foiled by Eve, so that they could not accurately show the power of light from the channel. For example, a short pulse light might reduce the average power of Eve's light, and the finite bandwidth of these monitor devices might worsen the monitoring results. Furthermore, a recent paper also shows other imperfections of a practical monitoring photodetector \cite{Sajeed14}.

\begin{figure}
\scalebox{1}{\includegraphics[width=\columnwidth]{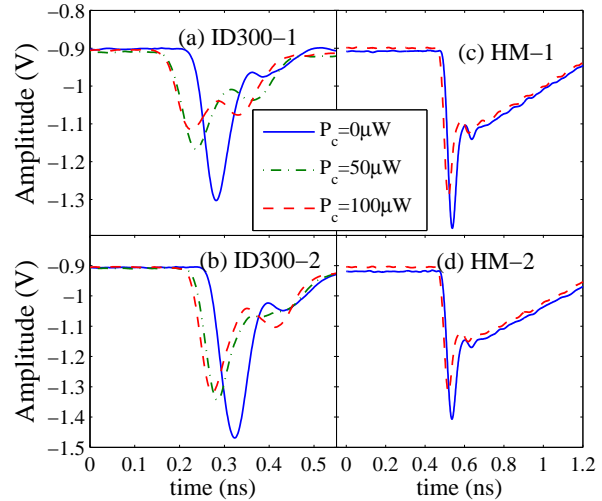}}
\caption{\label{fig:waveform}(Color online) Measured signal pulse waveforms when Alice's signal laser is illuminated by a bright light. Eve sends a bright cw light to Alice's signal lasers (including both commercial and homemade pulsed lasers), then the signal pulse of Alice is directly measured using a photodiode ($D_1$) with bandwidth 40GHz, an oscilloscope with bandwidth 33GHz and sample rate 80GHz (model: DSOX93303Q, Agilent). The repetition rate of the signal laser is 10MHz. It is clearly seen that when Alice's signal laser is illuminated, the pulse amplitude and width will be changed.}
\end{figure}

An active phase randomization (Fig.\ref{fig:attack}(c)) \cite{Zhao07}, or the cw laser followed by an external intensity modulator and an active phase randomization scheme, is another important choice for practical QKD systems, especially when the QKD system works in a high repetition rate \cite{Kobayashi14}. Then phase randomization assumption is automatically guaranteed. But such a countermeasure may not remove our attack entirely, since Eve can tamper with other parameters (e.g., intensity and shape, see Fig.\ref{fig:waveform}) to compromise the security of such systems. For example, the key rate of both CV-QKD and DV-QKD depends on the intensity of the signal pules \cite{Ma13,Wang-Peng08,Mizutani15}. But the stability of S-LD (no matter whether it works on pulsed mode or cw mode) could be damaged by bright light, so that the intensity of Alice's laser is unstable. Therefore, in this sense, our attack is also effective for the QKD system with a cw laser and an active phase randomization scheme. Another countermeasure is to use a protocol (or security proof) with an unrandom phase, but the performance of such a protocol is dramatically reduced in distance and key rate\cite{Lo07}.

\section{Discussion}
Fig.\ref{fig:waveform} shows that the pulse shape would also be changed by Eve's bright light. These changed parameters are also helpful. For example, the signal pulse is emitted earlier than that without Eve \cite{comm_timeshift}, and the time shift is different for each S-LDs. Furthermore, in the absence of an external field, the first oscillation is much stronger than the following oscillation, and a few oscillations appear \cite{comment_expfig6}. But when Eve is present, more oscillations are observed, and different laser diodes have different oscillation waveform. Thus it is possible for Eve to compromise the security of QKD systems with multi-lasers \cite{Schmitt07} by measuring the characters of signal pules (e.g., time-shift, pulse width, optical frequency).

Here we remark that, generally speaking, the changes of pulse shape are helpful for both Eve and Alice. Although more imperfection could be exploited by Eve, more parameters could be monitored by Alice to discover the existence of Eve. In fact, both Eve and Alice must be very careful in the cat-and-mouse game (see Appendix \ref{appendix_b} for details).  First, if Alice wants to completely monitor the changes of pulse shape, some advanced devices with high speed and bandwidth are required, which may dramatically increase the technology challenge and cost of a practical Alice. Second, Eve could carefully configure her attack to ensure that her attack could not increase the error rate and the changes of pulse shape could not be discovered by Alice. Third, generally speaking, the changed shape may actually benefit Eve more than Alice and Bob. This is because Eve could well be a spy or national security agency such as the NSA and so Eve has a much larger power and budget than Alice and Bob. Thus Eve is probably at a better position to exploit the imperfections that she has introduced in the quantum signal. Furthermore, note that even a tiny violation of the phase randomization assumption or other parameters of the source will undermine the very foundation of security proofs in QKD and it will no longer be fair for Alice and Bob to claim unconditional security.

Finally, in addition to using a laser, Eve can also attack the QKD system by using temperature, microwave radiation, and so on. At the same time, although most quantum hackers focus on the optical devices of the legitimate parties, Eve can also exploit imperfections in the electrical devices of the QKD system. For example, if the electromagnetic shielding of devices of Alice and Bob is imperfect, Eve could use microwave radiation from outside to control the parameters of these devices.  These are the subjects of future investigations.

\section{Conclusion}
In summary, phase randomization is a cornerstone assumption for many quantum communication protocols, and a tiny violation of such an assumption is fatal to the security of such protocols. However, here we demonstrate experimentally, with both commercial and homemade pulsed lasers, how easy it is for Eve to violate such a fundamental assumption in a practical setting. Additionally, besides the random phase, other parameters (e.g., intensity) of the source could also be changed. Our attack works for most DV-QKD protocols, and possibly for CV-QKD and other quantum information processing protocols (e.g., QCT and BQC).  Thus our work constitutes an important step towards secure quantum information processing.

\section{Acknowledgement}
We thank Z. Yuan and V. Makarov for helpful discussions. This work is supported by the National Natural Science Foundation of China, Grant No. 11304391. L.M.L is supported by the NCET program. H.-K. Lo is supported by NSERC. F. Xu is supported by the Office of Naval Research (ONR) and the Air Force Office of Scientific Research (AFOSR).

\appendix
\section{The scheme for Eve to foil Alice's monitor devices}\label{appendix_a}
Now we show that Alice's countermeasure, shown in Fig.\ref{fig:attack}(b), of the main text (include an isolator, an optical filter, and a photodetector), can't remove our attack entirely.

\emph{(i) Isolator-} In general, an optical isolator serves to prevent back-reflected photons from returning to Alice's lab. However, owing to the finite isolation of practical isolators, this approach only reduces the probability that photons infuse into Alice's zone, but can not eliminate this probability entirely. We perform a proof-of-principle experiment, by inserting a 25dB isolator after the output port of the signal laser ID300-1. The experimental results of Fig.\ref{fig:pro_isolator} of the main text show that the intensity distribution is still Gaussian-type but not U-type when Eve uses a cw laser with a power of $0.6mW$. Thus the phase of adjacent pulses can be still correlated. Although isolation of some commercial isolators reaches 50dB (or Alice can use two or more isolators in series to increase the isolation), it can not totally foil our attack, because Eve can always increase the power of her control laser. Furthermore, other imperfections of the practical isolator have been found in a recent paper \cite{Jain14}.

\emph{(ii) Filter-} An optical frequency filter is often used by Alice to remove any wavelength-dependent flaws. By doing so, only the light within a narrow band of frequencies can enter Alice's lab. However, the wavelength of Eve's control laser is the same as that of Alice's signal laser in our attack. Thus, an optical frequency filter is not an effective countermeasure against our attack.

\emph{(iii) Photodetector-} Alice can use both an optical power meter and photodetector to monitor the intensity of light from a quantum channel, but the optical power meter measures the average power of light. Thus it could be foiled by Eve who uses a pulsed laser. For example, Fig.4 of the main text shows that a cw laser with an optical power $0.6mW$ is sufficient to correlate the phase of Alice's signal pulse. Now, suppose that the repetition rate of the QKD system is 10MHz, and Eve uses a pulsed control laser with width of 100ps. Then the duty circle of Eve's pulse is 100ps/10ns=0.001. Thus the average optical power is reduced to $0.6mW \times 0.001= 0.6\mu W$.

A classical photodetector with a discrimination voltage can be used to monitor the intensity of pulsed light. However, the classical photodetector could also be cheated due to the following two reasons.

First, the classical photodetector can be damaged by bright light so that it may not work as expected. There are two kinds of classical photodetectors: one based on the PIN, and the other one based on the APD. Both can be damaged by bright light \cite{Bugge14}. For example, the detector based on InGaAs-APD from Thorlabs has a maximal input power 10mW (model: APD310) and 1mW (model: APD110C). The maximal input power for the detector based on InGaAs-PIN from Thorlabs (model: PDA8GS) is about 1mW for cw and 20mW for 60ms \cite{thorlabs}.

\begin{figure}
\scalebox{1}{\includegraphics[width=\columnwidth]{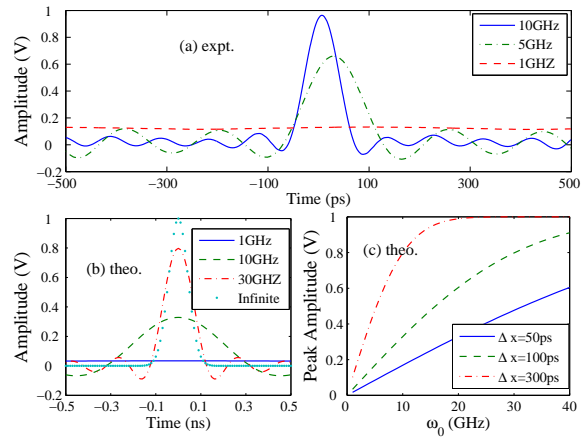}}
\caption{\label{fig:pulsewidth}(Color online) Part(a) shows the experimentally measured pulse amplitudes by directly inputting an electrical signal with amplitude 1V and 3dB width 100ps into an oscilloscope (model: DSOX93304Q, Agilent) with various bandwidth 1GHz, 5GHz, and 10GHz. The electrical signal is generated from a Patter Generator (model: 12050, Picosecond). Part(b) and Part(c) show the theoretical amplitude of an ideal Gaussian pulse which passes an ideal low-pass filter. $\Delta x$ is the 3dB width of the Gaussian signal. $\omega_0$ is the maximal bandwidth of the low-pass filter. In part(b), we set $\Delta x=100ps$. The mismatch between experiment (part(a)) and theory (part(b)) is mainly due to the simplified version of our model. All the results show that the monitor devices with finite bandwidth can not faithfully characterize the input signal.}
\end{figure}

Second, the finite bandwidth of the classical photodetector may worsen the monitoring results. We experimentally measure the amplitude of an electrical signal using an oscilloscope with various bandwidth (Fig.\ref{fig:pulsewidth}(a)). Furthermore, the theoretical amplitudes of an ideal Gaussian pulse which passes a linear time-invariant ideal low-pass filter are also shown in Fig.\ref{fig:pulsewidth}(b)-(c). Generally speaking, when a signal pulse, $f(t)$, passes a linear time-invariant device, its amplitude function becomes
\begin{equation}\label{eq_gt}
g(t)=\int_{-\infty}^{\infty} G(\omega) F[f(t)]e^{i\omega t}d\omega,
\end{equation}
where $F[\cdot]$ is the Fourier transformation, and $G(\omega)$ is the frequency response function of device.
It clearly shows that devices with finite bandwidth will filter high-frequency signals, and reduce the amplitude of a signal pulse. For simply, we assume that the signal is Gaussian pulse and the device is an ideal low-pass filter, that is,
\begin{equation}\label{eq_con}
\begin{split}
f(t)&=\exp[-\frac{t^2}{2\sigma^2}],\\
G(\omega)&=\begin{cases} 1& |\omega|\leq\omega_0\\ 0& |\omega| >\omega_0\end{cases}.
\end{split}
\end{equation}
Here $\sigma$ is the standard deviation of a signal pulse $f(x)$. If the 3dB width of $f(x)$ is noted as $\Delta x$, it is easy to check that $\Delta x=\sqrt{8\ln(2)}\sigma$. $\omega_0$ is the maximal bandwidth of the ideal low-pass filter.

The theoretical amplitude of $g(t)$ is shown in Fig.5(b)-(c) of the main text. The results clearly show that monitoring devices with finite bandwidth could not faithfully characterize the factual amplitude of the input signal, and Eve could foil the monitoring devices with a sharp pulsed signal. Although the test of Fig.\ref{fig:pulsewidth} is performed for an electrical signal, the results can be directly applied to the photodetector with finite bandwidth. For example, suppose that the gain and discrimination voltage of the photodetector are $10^4$ V/W and 0.2 V, and that Eve uses a pulsed control light with a 3dB width of 100ps and a peak power of $100\mu W$. Then the expected output voltage of the photodetector should be 1V, which is much larger than the discrimination voltage, 0.2V.

Fig.\ref{fig:waveform} of the main text also shows that if the bandwidth of Alice's photodetector is high enough (e.g., $>5GHz$), Eve can be discovered. (Note that generally speaking, the gain of photodetector will be decreased when the bandwidth is increased. But here we simply assume the gain is independent of the bandwidth.) However, if the bandwidth of photodetector is limited (e.g., 1GHz), the factual output voltage is lower than the discrimination voltage, 0.2V. Alice can not discover the existence of Eve. Note that Fig.\ref{fig:pro} of the main text has shown that $100\mu W$ is sufficient for Eve to break the phase randomization assumption. Furthermore, a recent paper also shows other imperfections in a practical monitoring photodetector \cite{Sajeed14}.

Therefore, the possible countermeasure of Fig.\ref{fig:attack}(b) of the main text could be cheated by Eve, if the devices are not carefully configured. Furthermore, illumination by a bright light changes not only the phase but also the pulse waveform, including its width, amplitude, and shape. Although we still do not know how Eve can obtain more information by exploiting such a modified waveform, it remains possible for Eve to attack the QKD system.

\section{A simple discussion about Fig.\ref{fig:waveform}}\label{appendix_b}
Fig.\ref{fig:waveform} of the main text clearly shows that when the signal laser is illuminated by bright light, the pulse shape would also be changed. Generally speaking, the additional changes are helpful for both Eve and the legitimate parties. More imperfections can be exploited by Eve to spy the final key, and more parameters can be monitored by Alice to discover the existence of Eve.  But it is still possible for Eve to perform our attack.

Theoretically speaking, Eve could perform a suitable attack to ensure that the modification of the pulse shape would not increase the error rate between Alice and Bob. In fact, Eve can perform the intercept-and-resend attack, and ensure that the error rate is lower than a reasonable value. For example, in the system with multi-laser diodes, she first measures the time-shift of each laser diode to determine Alice's state. Then she can resend a faked state to Bob according to her measurement results. In this case, if the time-shift is distinguishable for each laser diode (it is possible according to Fig.\ref{fig:waveform}), Eve could know the state sent by Alice. Then she can resend a perfect faked state to Bob according to her measurement. Thus no additional error will be introduced, and the legitimate parties could not discover the existence of Eve by monitoring the error rate.

Therefore, the main battlefield for Alice and Eve is the monitor devices, and both of them must be very careful in the cat-and-mouse game.

For Alice, she may discover the existence of Eve by carefully monitoring the parameters of the signal laser. But since the change is tiny in some parameters, some advanced devices with high speed and bandwidth  (e.g., photodetectors, analog-digital convertors, or time-amplitude convertors, and so on) are required for Alice, which may dramatically increase the technology challenge and cost of a practical Alice.  For example, the time-shift for ID300 lasers is about 100ps; thus if Alice wanted to characterize the time-shift of her pulses, the bandwidth and sample rate of Alice's analog-digital convertor should be larger than 40GHz (generally speaking, at least four points are needed to recover a pulse).  Furthermore, the bandwidth and sample rate should be increased for homemade lasers (see Fig.\ref{fig:waveform} of the main text for HM-1 and HM-2), since much smaller changes are introduced.

For Eve's part, she should carefully configure her attack to foil Alice's monitor devices. (1) Eve may carefully stable her controlling laser, and match the optical frequency of her controlling laser with that of Alice's signal laser, so that, excepting the random phase, many tiny changes will be introduced on the pulse shape. Taking the homemade lasers (HM-1 and HM-2) as an example, Eve's light will correlate the phase of each of the pulses (see Fig.\ref{fig:pro} (c) and (d) of the main text), but Fig.\ref{fig:waveform} (c) and (d) of the main text show that the changes of pulse shape are very tiny (At least, compared with ID300-1 and ID300-2, we do not find any obvious changes in the pulse shape using a photodetector with 40GHz bandwidth, an oscilloscope with 33GHz bandwidth and a sample rate of 80GHz, thus if Alice wants to discover the changed shape of HM-1 and HM-2, advanced devices with higher bandwidth and sample rate are required. (2) Eve may reduce the risk of being discovered by spying parts (not all) of final key.  For example, it has been proven that a small fluctuation of intensity will dramatically reduce the secret key rate of decoy state BB84 protocol \cite{Wang09}. Thus she still could obtain parts of final key by trivially changing the intensity of Alice's signal laser. In fact, it had been shown that, if the intensity of Alice's signal pulses fluctuates 1\%, 2\% and 3\%, the final key rate will be reduced by 11.86\%, 23.91\% and 36.17\% \cite{Wang09} (The simulation was performed based on the experimental parameters of Ref.\cite{Schmitt07}(b) ).

Furthermore, generally speaking, the ability for Eve to change other parameters in an optical signal may actually benefit Eve more than Alice and Bob. This is because Eve could well be a spy or work for a national security agency such as the NSA, and so Eve has a much larger budget than Alice and Bob, and thus is probably in a better position to exploit the imperfections that she has introduced in the quantum signal.



\begin{thebibliography}{100}
\bibitem{comm_timeshift} The time shift appears because when the laser cavity is seeded with an external field, the relaxation oscillations are dampened and the first oscillation occurs earlier. And the amount of the time shift is much larger for the two ID300 lasers than those of the homemade lasers, because the linewidth of two ID300 lasers is larger than that of the homemade lasers. Thus the probability that a photon is infused into the ID300 lasers is larger than that of the homemade lasers.
\end{thebibliography}
\end{document}